\documentstyle[preprint,aps]{revtex}
\tightenlines

\begin{document}
\newcommand{\beq}{\begin{equation}}
\newcommand{\eeq}{\end{equation}}
\newcommand{\beqa}{\begin{eqnarray}}
\newcommand{\eeqa}{\end{eqnarray}}
\newcommand{\fr}{\frac}
\draft
\preprint{INJE-TP-01-08, hep-th/0110123}

\title{Entropy of the three dimensional Schwarzschild-de Sitter black hole }

\author{ Y.S. Myung\footnote{Email-address :
ysmyung@physics.inje.ac.kr}}
\address{Relativity Research Center and School of Computer Aided Science, Inje University, Kimhae 621-749, Korea}

\maketitle

\begin{abstract}
We study the three dimensional Schwarzschild-de Sitter (SdS$_3$) black hole
which corresponds essentially  to a conical defect.
We compute the  mass of the SdS$_3$ black hole from the correct
definition of the mass in asymptotically de Sitter space.
Then we clarify  the relation between the mass, entropy and temperature for
this black hole without any ambiguity.
Also we establish the SdS$_3$/CFT$_2$-correspondence for the
entropy by applying the Cardy formula to a CFT with a central
charge $c=3\ell/2G_3$. Finally we discuss the entropy bounds for
the SdS$_3$ black hole.

\end{abstract}
\vfill
Compiled at \today : \number \time.

\newpage

Recently an accelerating universe has proposed to be a way
to interpret the astronomical data of supernova\cite{Per,CDS,Gar}.
Combining this observation with the need of inflation  in the
standard cosmology leads to that our universe approaches de Sitter
geometries in both the far past and the far future\cite{Wit,HKS,FKMP}. Hence it is
very important to study the nature of de Sitter space. However,
there are two  difficulties in studying de Sitter space :
First there is no spatial infinity and second there is no global timelike
Killing vector.  Thus it is not easy to defin the conserved  quantities including  mass,
charge and angular momentum appeared in the  de Sitter black holes.

More recently authors in\cite{SSV} proposed the mass (${\cal M}$),
temperature ($T$)
and entropy ($S$) for the SdS$_3$ black hole. But there is some
ambiguity in defining the mass  (or energy ) of this black hole.
They
used a relation of $dS/d(-{\cal M})=1/T$ to derive the
entropy. We wish to point out that although this provides us the entropy, this is not a correct
relation between these quantities. If one defines the mass ($M$) of this black hole properly,
one can use  the correct relation of $dS/dM=1/T$(
the first law of the SdS$_3$ black hole thermodynamics) to find the entropy.
Furthermore one guesses the SdS/CFT correspondence from
the dS/CFT correspondence\cite{STR,Wit}, as
defined analogously to the AdS/CFT correspondence\cite{MAL}. So it is
important to establish this correspondence by computing both the
boundary CFT entropy and bulk SdS entropy.
For this purpose, the SdS$_3$ black hole plays the  role of a toy model.
 Also there exist the entropy bounds of  $N$-bound and $D$-bound
 for objects in asymptotically de Sitter
space and Bekenstein($B$)-bound for asymptotically flat space\cite{BOU}.
 Hence it is interesting to study the relation between these bounds for the
SdS$_3$ black hole (a pointlike object) itself.

In this letter,
we calculate the  mass of the SdS$_3$ black hole using the correct
definition of the mass in asymptotically de Sitter space\cite{BBM}.
Then we clarify  the relation between the mass, entropy and temperature
 without any ambiguity.
Also we establish the SdS$_3$/CFT$_2$-correspondence for the
entropy by applying the Cardy formula to the CFT$-2$ with a central
charge $c=3\ell/2G_3$. Finally we discuss the entropy bounds for
the SdS$_3$ black hole.

We start with the d-dimensional Schwarzschild-de Sitter metric in the static
coordinates\cite{SSV}
\beq
ds_d^2=-\Big(1-\fr{2m}{r^{d-3}}-\fr{r^2}{\ell^2}\Big)dt^2 +
\Big(1-\fr{2m}{r^{d-3}}-\fr{r^2}{\ell^2}\Big)^{-1}dr^2 +r^2
d\Omega^2_{d-2},
\eeq
where $2m= 16 \pi G_d {\cal M}/Vol(S^{d-2})$ is a parameter related to the black hole mass
 \cite{CMO} and $\ell$
is the curvature radius of de Sitter space. This is solution to
the action

\beq
S_{d} = {1 \over 16 \pi G_d} \int d^d x \sqrt{-g} \left [
 R -2 \Lambda \right]
\label{Qact}
\eeq
with the d-dimensional Newtonian   constant $G_d$ and the
d-dimensional positive cosmological constant
$\Lambda=(d-1)(d-2)/2\ell^2$.

For the three dimensional Schwarzschild-de Sitter black hole, we have
 the metric as
\beq
ds_3^2=-\Big(1-8G_3 {\cal M}-\fr{r^2}{\ell^2}\Big)dt^2 +
\Big(1-8G_3 {\cal M}-\fr{r^2}{\ell^2}\Big
)^{-1}dr^2 +r^2
d\phi^2
\label{met-0}
\eeq
which can be rewritten by the new parameter
$r_c=\sqrt{1-8G_3 {\cal M}}$ as

\beq
ds_3^2=-\Big(r_c^2-\fr{r^2}{\ell^2}\Big)dt^2 +
\Big(r_c^2-\fr{r^2}{\ell^2}\Big)^{-1}dr^2 +r^2
d\phi^2.
\label{met-1}
\eeq
In the case of $r_c=1 ({\cal M}=0)$, we have de Sitter(dS$_3$)
space. Thus we may identify ${\cal M}$ as the mass of the SdS$_3$ black hole.
However, this attempt leads to a wrong result.
We note here that $0<r_c < 1$ if ${\cal M} \not=0$.
Hence this describes a conical defect spacetime with deficit angle
$2\pi(1-r_c)$, indicating a world with a positive cosmological
constant and a pointlike massive object with ${\cal M}$\cite{SSV}.
 In oder to obtain a conserved quantity of the  mass ($M$),
we have to use a proper definition of the mass in an asymptotically de
Sitter space as\cite{BBM}
\beq
M= \oint_{\Sigma} d\phi \sqrt{\sigma} N_\rho n^\mu n^\nu T_{\mu\nu}.
\eeq

We use this formula on a surface of fixed time and then
send time to infinity so it approaches the past(future) infinity
at ${\cal I^-}({\cal I^+})$. This is so because in a theory of
gravity, mass is measure of how much a metric deviates near
infinity from its vacuum state. In order to evaluate mass
properly, we need an equal time surface $\Sigma$ of the conical defect
spacetime outside the cosmological horizon ($r>\ell r_c$) with
\beq
ds_{\Sigma}^2=(\fr{r^2}{\ell^2}-r_c^2)dt^2 + r^2
d\theta^2.
\eeq
It is
noted that for $r<\ell r_c$, $t(r)$ is timelike (spacelike) while $r>\ell r_c$, $t(r)$ is
spacelike (timelike).
As $r \to \infty$, we thus have

\beq
M= \fr{1}{8\pi G_3}\oint_{{\cal I}^{\pm}}d\theta \fr{r_c^2}{2}=\fr{r_c^2}{8G_3}=\fr{1}{8G_3}
-{\cal M}.
\eeq
This defines the conserved quantity of mass (energy :$E$) for the SdS$_3$
black hole. If ${\cal M}=0$,
we recover a mass $M_{dS}=1/8G_3$ of
the pure de Sitter space with a cosmological horizon at $r_0=\ell$.
If ${\cal M}\not=0$, we find a mass of the SdS$_3$ black hole
 which is less than that
 of pure de Sitter space $M_{dS}$. Then we ask of why a matter ${\cal M}$
  contributes
 to $M$ as a negative one.  The answer is that
even if the matter of a pontlike object has a positive energy, the binding energy to the
gravitational de Sitter background can make this a negative one.
As a result, the quantity ${\cal M}$ is not considered as  the true
mass (energy) of the SdS$_3$
black hole. Authors in\cite{SSV} used  this mass ${\cal M}$
 to derive the entropy of the SdS$_3$
black hole using  $dS/d(-{\cal M})=1/T$. Although this provides the entropy,
 this is not a correct
expression for the first law of the black hole thermodynamics.

Now we have to use  the correct expression to obtain the entropy.
The temperature $T$ of this black hole can be easily found from the
fact that considering Eq.(\ref{met-0}), the Euclidean green function is periodic in imaginary
time ($\tau)$ with periodicity
\beq
\tau \to \tau + i \beta, \beta=\fr{2 \pi \ell}{r_c}.
\eeq
From this we find the Hawking temperature
\beq
T=\fr{r_c}{2 \pi \ell}.
\eeq
Here we consider $M,T,S$ as functions of $r_c$ : $M=M(r_c);
T=T(r_c); S=S(r_c)$.
Using the correct relation of the black hole thermodynamics
\beq
dS/dM=1/T,
\eeq
we obtain the Bekenstein-Hawking entropy

\beq
S_{SdS}=\fr{\pi \ell r_c}{2G_3}.
\label{BEN}
\eeq
The bulk entropy $S_{dS}=\fr{\pi \ell}{2G_3}$ with $r_c=1$ for
the dS$_3$ space appeared in\cite{BBO}. We can easily find that
for $0<r_c<1$, $S_{SdS}<S_{dS}$.

Also the  dS$_3$/CFT$_2$ correspondence was proposed by using
the AdS$_3$/CFT$_2$ correspondence\cite{STR,Wit,MAL}.
Now we are in a position to discuss the SdS$_3$/CFT$_2$
correspondence. This can be regarded as an extended version of the
dS$_3$/CFT$_2$ correspondence. For this purpose,
we use the Cardy formula together with the central
charge. From the anomalous transformation
law of the stress-energy tensor $ T_{++}=-\fr{\ell}{16 \pi G_3} \partial_+^3 \xi^+
(T_{--}=-\fr{\ell}{16 \pi G_3} \partial_-^3 \xi^-)$ in a two-dimensional conformal
theory at ${\cal I}^{\pm}$, we can read off the central charge
$c$\cite{STR,BBM}
\beq
c=\fr{3\ell}{2G_3}.
\eeq
The eigenvales of the conformal generators $L_0$ and $\bar L_0$ in
static coordinates are given by\cite{BBM}

\beq
L_0=\bar L_0=\fr{M\ell}{2}.
\eeq
The Cardy formula for the asymptotic density of states of a
unitary CFT is
\beq
S_{CFT}= 2\pi \sqrt {\fr{c L_0}{6}} + 2\pi \sqrt {\fr{c \bar L_0}{6}}
\eeq
which gives us the boundary entropy at ${\cal I}^{\pm}$.
This exactly leads to the bulk entropy Eq.(\ref{BEN}) for the  SdS$_3$ black
hole as
\beq
S_{CFT}= 4 \pi \ell \sqrt{\fr{M}{8G_3}}=\fr{\pi \ell
r_c}{2G_3}=S_{SdS}.
\eeq
This establish the SdS$_3$/CFT$_2$
correspondence for the entropy.

Finally we discuss the entropy bounds for the SdS$_3$ black hole
(a pointlike object with ${\cal M})$
which is located at $r=0$\cite{BOU,CMO}. This corresponds to a very small,
dilute object in asymptotically de Sitter space.
The $N$-bound means that the entropy of an object in
asymptotically de Sitter space is bounded by the de Sitter entropy
: $S_{{\cal M}}^N \le S_{dS}$. Further, applying the Geroch
process to the cosmological horizon $r=r_c \ell $ leads to the
entropy of an object in asymptotically de Sitter space is bounded by the
difference of the entropies in  pure de Sitter space and in asymptotically de Sitter
space.
This is called $D$-bound : $S_{{\cal M}}^D \le S_{dS}-S_{SdS}$.
In addition, for a system of volume $V$ (linear size $R$) with the limited
self-gravity, the total entropy of the system satisfies the
Bekenstein bound ($B$-bound) : $S_{{\cal M}}^D \le 2 \pi R {\cal E}$. Here ${\cal E}$
is the energy of the system. This  is the flat space bound and thus
has nothing intrinsically to do with the de Sitter bounds like $N$-bound and  $D$-bound.

Introducing the gravitational radius
 $r_g=8G_3 {\cal M}$ in Eq.(\ref{met-0}), this corresponds a
 well-defined quantity in asymptotically de Sitter space.
Here we have a relation $r_g=(r_0-\ell r_c)(r_0+\ell r_c)/r_0^2$.
 In this case of $r_g <<1$ (a very small and
dilute object localized at $r=0$),
we have $r_c \approx 1-4G_3 {\cal M}$.
Then $D$-bound reduces to $S_{{\cal M}}^D \le 2\pi r_0 {\cal M}$.
Replacing $r_0 \approx r_c$ by $R$ and ${\cal M} $ by ${\cal E}$,
$D$-bound leads to $B$-bound :$S_{{\cal M}}^D \to S_{{\cal M}}^B
\le 2 \pi R {\cal E}$. This shows an example for the relation between
$D$-bound and $B$-bound : for dilute spherical systems,
$D$-bound in asymptotically de Sitter space coincides with $B$-bound in flat space\cite{BOU}.

In conclusion, we calculate the  mass of the SdS$_3$ black hole using the correct
definition of the mass in asymptotically de Sitter space\cite{BBM}.
Then we use  the first law of thermodynamics which is the
relation between the mass, entropy and temperature
to obtain the bulk Bekenstein-Hawking entropy.
Also we establish the SdS$_3$/CFT$_2$-correspondence for the
entropy by applying the Cardy formula to the CFT$_2$ with a central
charge $c=3\ell/2G_3$. Finally  we show that
$D$-bound coincides with $B$-bound in flat space.

\section*{Acknowledgement}
We thank  H.W. Lee for helpful discussions.
This work was supported in part by the Brain Korea 21
Program of  Ministry of Education, Project No. D-1123 and
 KOSEF, Project No. 2000-1-11200-001-3.

\end{document}